\documentclass[twocolumn,showpacs,preprintnumbers,superscriptaddress,aps]{revtex4-1}

\usepackage{bbm, bm, epsfig, amsmath, amsfonts, amssymb, wasysym, graphicx, color}
\usepackage[colorlinks,linkcolor=blue,urlcolor=black,citecolor=blue]{hyperref}
\usepackage{natbib}

\definecolor{darkgreen}{rgb}{0,0.5,0}
\definecolor{purple}{rgb}{0.35,0,0.35}
\definecolor{orange}{rgb}{1,0.5,0}
\definecolor{darkred}{rgb}{.7,0,0}
\definecolor{darkblue}{rgb}{0.1,0.1,.6}
\definecolor{blue}{rgb}{0,0,.8}
\definecolor{grey}{rgb}{.6,.6,.6}
\definecolor{dimgreen}{rgb}{0.2,0.6,0.1}
\definecolor{RoyalBlue}{rgb}{0,0.416,0.702}
\definecolor{DGLorange}{rgb}{0.949,0.573,0}

\newcommand{\bsigma}{\bar{\sigma}}

\newcommand{\BSOI}{{\bm{B}_{\rm SOI}}}
\newcommand{\B}{{\bm{B}}}

\newcommand{\Vg}{V_{\rm g}}
\newcommand{\GQ}{G_{\rm Q}}
\newcommand{\Omegax}{\Omega_x}
\newcommand{\Omegaxone}{\Omega_{x1}}
\newcommand{\Omegaxtwo}{\Omega_{x2}}

\newcommand{\Cbarrier}{C_{\rm b}}
\newcommand{\Cdisp}{C_{\rm d}}

\newcommand{\ddL}{\frac{\textrm{d}}{\textrm{d} \Lambda}}
\newcommand{\gammaone}{\gamma_1^\Lambda}
\newcommand{\gammatwo}{\gamma_2^\Lambda}
\newcommand{\gammap}{\gamma_p^\Lambda}
\newcommand{\gammax}{\gamma_x^\Lambda}
\newcommand{\gammad}{\gamma_d^\Lambda}

\begin{document}

\relpenalty=9999
\binoppenalty=9999

\title{Effect of spin-orbit interactions on the 0.7 anomaly in quantum point contacts}

\author{Olga Goulko}
\affiliation{Physics Department, Arnold Sommerfeld Center for Theoretical Physics, and Center for NanoScience,
Ludwig-Maximilians-Universit\"at, Theresienstra\ss e 37, 80333 Munich, Germany}
\affiliation{Present address: Department of Physics, University of Massachusetts, Amherst, MA 01003, USA}
\author{Florian Bauer}
\affiliation{Physics Department, Arnold Sommerfeld Center for Theoretical Physics, and Center for NanoScience,
Ludwig-Maximilians-Universit\"at, Theresienstra\ss e 37, 80333 Munich, Germany}
\author{Jan Heyder}
\affiliation{Physics Department, Arnold Sommerfeld Center for Theoretical Physics, and Center for NanoScience,
Ludwig-Maximilians-Universit\"at, Theresienstra\ss e 37, 80333 Munich, Germany}
\author{Jan von Delft}
\affiliation{Physics Department, Arnold Sommerfeld Center for Theoretical Physics, and Center for NanoScience,
Ludwig-Maximilians-Universit\"at, Theresienstra\ss e 37, 80333 Munich, Germany}

\begin{abstract}
  We study how the conductance of a quantum point contact is affected
  by spin-orbit interactions, for systems at zero temperature both
  with and without electron-electron interactions. In the presence of
  spin-orbit coupling, tuning the strength and direction of an
  external magnetic field can change the dispersion relation and hence
  the local density of states in the point contact region. This
  modifies the effect of electron-electron interactions, implying
  striking changes in the shape of the 0.7-anomaly and introducing
  additional distinctive features in the first conductance step.
\end{abstract}

\maketitle

Spin-orbit interactions (SOI) play an important role in a variety of
fields within mesoscopic physics, such as spintronics and topological
quantum systems. In this Letter we study the effects of SOI on the
conductance of a quantum point contact (QPC), a one-dimensional
constriction between two reservoirs \cite{wharametal,vanweesetal}. The
linear conductance $G$ of a QPC is quantized in multiples of
$\GQ=2e^2/h$, showing the famous staircase as a function of gate
voltage. In addition, at the onset of the first plateau, measured
curves show a shoulderlike structure near $0.7\GQ$
\cite{Thomas1996}. In this regime QPCs exhibit anomalous behavior in
the electrical and thermal conductance, noise, and thermopower
\cite{Thomas1996, Appleyard2000, Kristensen2000, Cronenwett2002,
  DiCarlo2006, Chiatti2006, Smith2011, Micolich2011, Danneau2008}. The
microscopic origin of this $0.7$-anomaly has been the subject of a
long debate \cite{Reilly2002, Berggren2002, Meir2002, Sloggett2008,
  Lunde2009, Aryanpour2009, BauerNature}. It has recently been
attributed to a strong enhancement of the effects of electron-electron
interactions (EEI) by a smeared van Hove singularity in the local
density of states (LDOS) at the bottom of the lowest QPC subband
\cite{Sloggett2008, BauerNature}. While this explains the
$0.7$-anomaly without evoking SOI, the presence of SOI can change the
dispersion relation and hence the LDOS, thus strongly affecting the
shape of the $0.7$-anomaly. Previous studies of SOI in QPCs exist
\cite{Kohda2012, Hsiao2010, Nowak2014, Danneau2006, Nichele2014}, but
not with the present emphasis on their interplay with the
  QPC barrier shape and EEI, which are crucial for
understanding the effect of SOI on the $0.7$-anomaly.

\textit{Setup.} We consider a heterostructure forming a
two-dimensional electron system (2DES) in the $xy$-plane. Gate
voltages are used to define a smooth, symmetric
potential which splits the 2DES into two leads, connected by a short,
one-dimensional channel along the $x$-axis: the QPC
\cite{wharametal,vanweesetal}. The transition between the leads and the QPC is adiabatic. We also assume the confining
    potential in the transverse direction to be so steep that the
  subband spacing is much larger than all other energy scales relevant for transport, in particular those related to the
  magnetic field and SOI, and consider only transport in the first
subband, corresponding to the lowest transverse mode. This can be
described by a one-dimensional model with a smooth potential barrier and local EEI \cite{BauerNature}. The
magnetic field $\B$ is assumed to be in the $xy$-plane, acting as a
pure Zeeman field, without orbital effects.

A moving electron in an electric field can experience an effective
magnetic field $\BSOI$ proportional to its momentum $\hbar
k$. Depending on the origin of the electric field one distinguishes
between Rashba and Dresselhaus terms, the former resulting from the
gradient of the external potential, and the latter from the asymmetry
of the ionic lattice \cite{winkler}. To be able to rotate $\B$ through
any angle $\varphi$ w.r.t.\ $\BSOI$ we require that $\BSOI$ also lies
in the $xy$-plane. W.l.o.g.\ (see Supplement) we choose the $y$-axis
to be parallel to $\BSOI$, such that the SOI contribution to the
Hamiltonian is $-\alpha\sigma_yk$, where $\alpha$ characterizes the
strength of the (Rashba) SOI and $\sigma_y$ is a Pauli matrix
\cite{Meieretal2007}. We only consider the leading SOI contribution
proportional to $k$ and choose the spin quantization direction along
$\B$.

\begin{figure*}
\includegraphics[width=\textwidth]{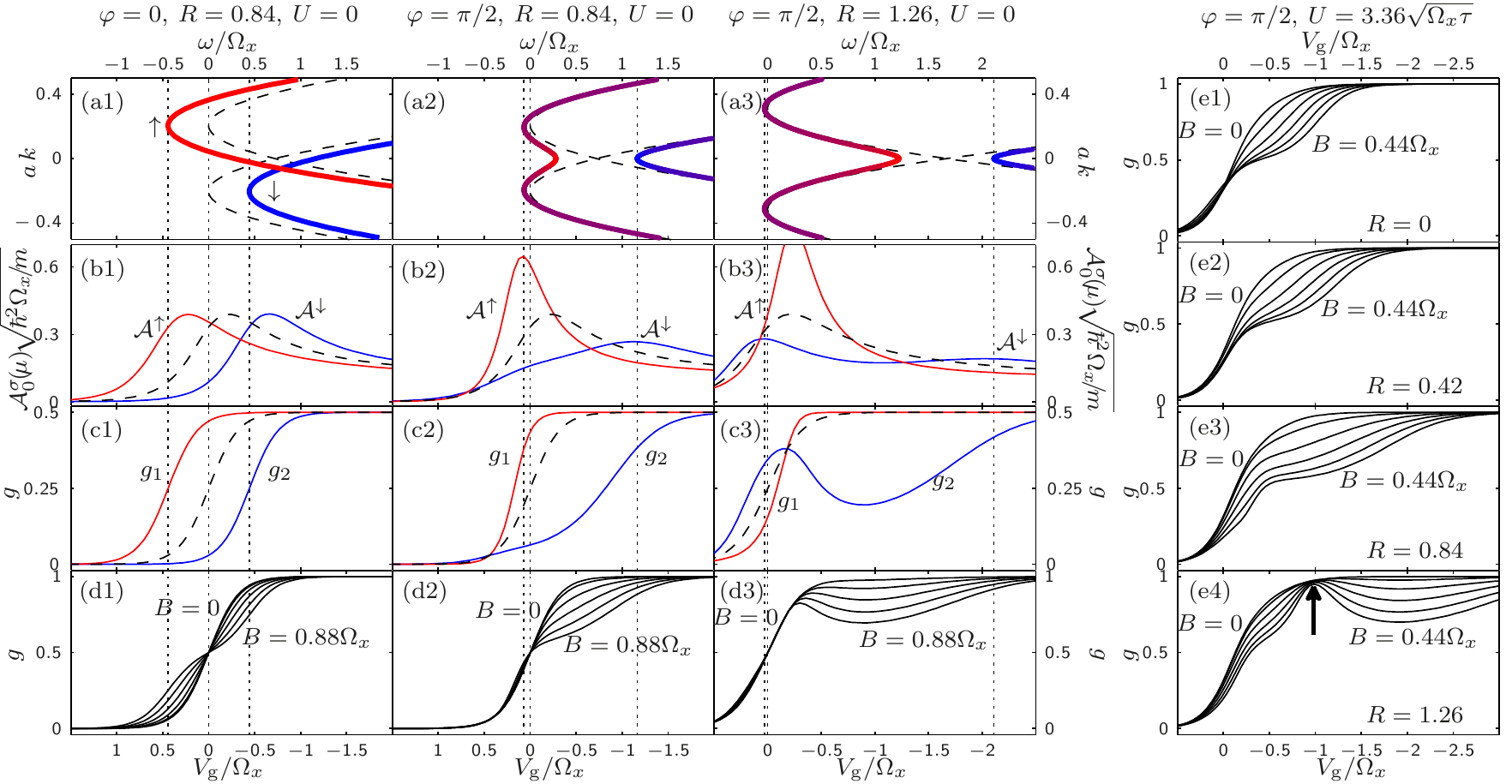}
\caption{\label{fig:disp}
Effect of SOI on the model without EEI, left columns (a1-d3), and with EEI, right column (e1-e4). The left columns (a1-d1), (a2-d2) and (a3-d3) represent different combinations of SOI strength $R$ and angle $\varphi$ between $\B$ and $\BSOI$. They highlight the correspondence between the dispersion relation $\omega(k)$ in a homogeneous system (a1-a3), the LDOS for fixed $\omega=\mu$ as function of $\Vg$ on the central site of a QPC with potential barrier (b1-b3), the conductances of the two QPC transmission channels (c1-c3), and the total conductance of the QPC for several equally-spaced magnetic field values between $B=0$ and $B=0.88\Omegax$ (d1-d3). In (a1-c3) the magnetic field is fixed at $B=0.88\Omegax$, with dashed lines showing the case $B=0$ for comparison. The line colors in (a1-a3) quantify the contribution of each spin state (red$=\uparrow$, blue$=\downarrow$) in the dispersion branches, to illustrate the spin mixing at $\varphi \neq 0$. The right column (e1-e4) shows the total conductance for $U > 0$, with $\varphi = \pi/2$ and several combinations of $R$ and $B$ (the latter were chosen smaller than in (d1-d3), since EEI enhance the $g$-factor \cite{BauerNature}).
}
\end{figure*}

Without SOI, the dispersion relation $\hbar^2k^2/2m$ of a homogeneous
one-dimensional model with effective mass $m$ splits in the presence
of a Zeeman field into two identical branches offset in energy by $\pm
B/2$. On the other hand, without a Zeeman field, the
momentum-dependent SOI splits the dispersion in $k$-direction and also
yields a negative spin-independent energy offset of magnitude $\Delta
E_{\rm SOI}=\alpha^2 m/2 \hbar^2$. In the following we shift the
energy origin by $-\Delta E_{\rm SOI}$ and quote all energies w.r.t.\
the new origin. If both $\B$ and $\BSOI$ are non-zero, their interplay
depends on $\varphi$, as illustrated in Fig.~\ref{fig:disp}(a1-a3). In
(a1), where the fields are parallel ($\varphi = 0$), the energy
offsets simply add, while for nonparallel fields a spin mixing occurs,
resulting in an avoided crossing \cite{Quayetal2010}. For orthogonal
fields ($\varphi = \pi/2$), the lower dispersion branch exhibits
either one broader minimum at $k=0$ if $B\geq4E_{\rm SOI}$, or two
minima at finite $k$ and a maximum at $k=0$ otherwise. The latter case
is shown in Fig.~\ref{fig:disp}(a2-a3).

\textit{Model. }For the lowest subband we model the QPC by a symmetric potential barrier which is quadratic around its maximum, 
\begin{equation}
V(x) \simeq \Vg + \mu - \Cbarrier x^2/2, 
\end{equation}
and vanishes smoothly at the boundary of the QPC. The barrier height $\Vg$, measured
w.r.t.\ the chemical potential $\mu$, mimics the role of the gate-voltage. If $\Vg$
is swept downwards through zero, the conductance $g=G/\GQ$ increases from 0 to 1. For
$B=0$ this occurs in a single step whose width is given by the
energy scale $\Omegax=\sqrt{\Cbarrier\Cdisp}$, which is set by the fixed curvature of the barrier, $\Cbarrier$,
and the curvature of the bulk dispersion at its minimum, $\Cdisp$ \cite{Buttiker1990}. For $\varphi=0$, $\Cdisp=\hbar^2/m$.

For numerical purposes we discretize real space and obtain an infinite tight-binding chain with spacing $a$, taking $\B$ and $\alpha$ constant throughout the chain. The noninteracting Hamiltonian is
\begin{eqnarray}
\label{eqModel}
H_0& =& \sum_{j,\sigma, \sigma'} d^\dagger_{j \sigma} \left[(V_j+2\tau) \delta_{\sigma \sigma'} - \frac{1}{2} (\pmb{\sigma}\cdot \B)_{\sigma\sigma'} \right] d_{j \sigma'}\\
\hspace*{-2mm}& &+\sum_{j,\sigma,\sigma'}\left[ d_{j+1 \sigma}^\dagger \left( -\tau_0 \delta_{\sigma \sigma'} + \frac{i\alpha}{2}(\sigma_y)_{\sigma\sigma'} \right) d_{j\sigma'}+ \textnormal{h.c.}\right]\!\!, \nonumber
\end{eqnarray}
where $d_{j\sigma}$ annihilates an electron with spin $\sigma$$\in$$\{\uparrow,\downarrow\}\equiv\{+,-\}$ at site $j$. The effective mass of the charge carrier is $m=\hbar^2/2\tau a^2$ with $\tau=\sqrt{\tau_0^2+\alpha^2}$ \cite{birkholzthesis}. We keep $\tau$ fixed when varying $\alpha$. The QPC barrier potential $V_j = V(ja)$ (and later EEI) are nonzero only in a region of length $L=2Na$ centered around $j=0$, representing the QPC. All results shown are for $N=50$. We use the smooth function $V(x)=(\Vg+\mu)\exp[-(2x/L)^2/(1-(2x/L)^2)]$ for the potential, with $\mu=2\tau$. Sites $j$$<$$-N$ and $j$$>$$N$ represent two leads with bandwidth $4\tau$. The strength of SOI in a QPC is determined by the dimensionless parameter
\begin{equation}
R=\sqrt{\frac{\Delta E_{\rm SOI}}{\Omega_x}}=\frac{\alpha}{\hbar}\sqrt{\frac{m}{2\Omega_x}}.
\end{equation}
SOI strengths of up to $\alpha\simeq10^{-11}$eVm have been reported in
the literature \cite{Nittaetal1997, Quayetal2010, Kohda2012,
  Kita2005}. Typical values of $\Omegax\simeq1$meV and
$m\simeq0.05m_e$ for InGaAs yield $R\simeq0.2$. A stronger spin-orbit
effect due to an enhancement of the anisotropic Lande $g$-factor is
reported in \cite{Martin2010}. Hole quantum wires have been used to
observe the spin-orbit gap \cite{Quayetal2010} and the anisotropic
Zeeman splitting \cite{Chen2010}. For hole QPCs, the larger effective
hole mass and the resulting smaller $\Omegax$ imply larger values of
$R$. Here we consider both small and large $R$, where $R\lesssim0.4$
is a realistic scale for electron systems and $R\gtrsim1$ is
accessible using hole systems \cite{prcomm2}, for QPCs with small
barrier curvature $\Cbarrier$ and hence small $\Omegax$. 

\textit{System without EEI. } Many insights on the interplay between
SOI and geometry can already be gained from the model without EEI, as
shown in the left part (a1-d3) of Fig.~\ref{fig:disp}. We discuss
exact results for two physical quantities, which we also relate to the
bulk dispersion relation: the linear conductance $g$ and the LDOS
$\mathcal{A}_j^{\sigma}(\omega)=-\textnormal{Im}\mathcal{G}^{\sigma\sigma}_{jj}(\omega)/\pi
a$, where $\mathcal{G}^{\sigma\sigma'}_{jj'}$ is the retarded
propagator from site $j'$ with spin $\sigma'$ to site $j$ with spin
$\sigma$. Due to SOI, spin is not conserved for $\varphi \neq 0$ and
hence $\mathcal{G}^{\sigma\sigma'}_{jj}$ is not spin-diagonal. However
at $j=0$ its off-diagonal elements turn out to be negligible compared
to the diagonal ones. Thus it is meaningful to analyze the LDOS at
$j=0$ for given $\sigma$. The linear conductance at zero temperature can be calculated via $g=g_1+g_2 \propto \textnormal{Tr}(t^\dagger t)$ \cite{datta}, where $t^{\sigma\sigma'} = \mathcal{G}^{\sigma\sigma'}_{-N,N}(\mu)$ is the transmission matrix of the QPC and $\textnormal{Tr}(t^\dagger t)$ equals the sum of the eigenvalues of $t^\dagger t$. The spin structure of $t$ depends on
$N$, but the eigenvalues of $t^\dagger t$, which yield the
conductances $g_1$ and $g_2$ of the two transmission channels, do not.

For $\varphi = 0$ (Fig.~\ref{fig:disp}, left column) spin is conserved and SOI have
no influence on the LDOS and the conductance. This case is analogous to the one
discussed in \cite{BauerNature}. The bulk (i.e.\ $V(x)=0$) LDOS,
\begin{equation}
\mathcal{A}^\sigma_{\rm bulk} (\omega)\propto\left.
\frac{\partial k}{\partial \omega}\right|_\sigma =\sqrt{\frac{m}{2\hbar^2(\omega +
\sigma B /2)}},
\end{equation}
has a van Hove singularity, diverging at the minimum $\omega$$=$$-\sigma B /2$ of the corresponding dispersion branch, where the electron velocity vanishes. In the QPC, the $x$-dependent LDOS is shifted in energy by the barrier potential $V(x)$. Since the barrier breaks translational invariance, the van Hove singularity is smeared out on a scale set by $\Omegax$ \cite{Sloggett2008}, forming a ridgelike structure, called van Hove-ridge in \cite{BauerNature}. The LDOS height becomes finite, of order $\mathcal{O}(\sqrt{m/(\hbar^2\Omegax)})$, determined by $\Omegax$ and the curvature $\hbar^2/m$ of the bulk dispersion. At a given position $x$, the LDOS maximum occurs at an energy which is $\mathcal{O}(\Omegax)$ larger than the corresponding potential energy $V(x)-\sigma B /2$. Here and henceforth we quote the LDOS as a function of $\Vg$ at fixed $\omega = \mu$. Figure~\ref{fig:disp}(b1) shows it at the central site $j = 0$; the spatially resolved LDOS is shown in Fig.~1 of the Supplement. The LDOS has the same shape for both spins. Its structure is clearly inherited from that of the dispersion in (a1), with peak energies aligned with the dispersion minima up to the shift of $\mathcal{O}(\Omegax)$. Similarly, the conductances $g_1(\Vg)$ and $g_2(\Vg)$ of the two channels in (c1) show steps of the same shape with widths $\propto\!\Omegax$ \cite{Buttiker1990}, split by $B$ and aligned with the dispersion minima. This causes the total conductance $g(\Vg)$ in (d1) to split symmetrically into a double step with increasing field, just as for a QPC without SOI.

Next consider the case $\varphi = \pi/2$ shown in
Fig.~\ref{fig:disp}(a2-d3). Spin mixing leads to an avoided crossing
with spin gap $\propto\!\!B$, which splits the dispersion into an
upper branch with a narrow minimum and a lower branch with two minima
and one maximum (for $B<4E_{\rm SOI}$). Note that bulk LDOS structures
separated in energy by less than $\Omegax$ are not resolved within the
QPC. In the following we give an intuitive explanation of how the
dispersion minima relate to the properties of the LDOS peaks and the
conductance steps. The curvatures of the lower and upper dispersion
branches are, respectively, smaller or larger than in (a1), $C_{\rm
  d1} < \Cdisp < C_{\rm d2}$ (loosely speaking, $C_{\rm d1}$ is the
effective curvature obtained by smearing the double dispersion minimum
by $\Omegax$, yielding a single minimum). Because the barrier
curvature $\Cbarrier$ is fixed, this results in two modified energy
scales $\Omega_{xi}=\sqrt{\Cbarrier C_{{\rm d}i}}$, with $\Omegaxone <
\Omegax < \Omegaxtwo$, which determine the LDOS peak heights and
widths, as well as the conductance step widths. Consequently, in (b2)
the LDOS peak for $\mathcal{A}^\downarrow_0$ is lower and wider than
for $\mathcal{A}^\uparrow_0$. Likewise, in (c2) the conductance step
for $g_2(\Vg)$ is wider than for $g_1(\Vg)$, causing $g(\Vg)$ in (d2)
to show a striking asymmetry for its $B$-induced evolution from a
single to a double step. This asymmetry is reminiscent of but
unrelated to that known for the 0.7-anomaly -- the latter is driven by
EEI, as discussed below -- but should be observable in higher
conductance steps, where EEI are weaker.

For $R\gtrsim1$ more structures emerge, see
Fig.~\ref{fig:disp}(a3-b3). Spin-mixing produces an additional
``emergent'' peak in $\mathcal{A}_0 ^ \downarrow$ (b3) and an
additional step in $g_2(\Vg)$ (c3) near $\Vg \simeq 0$. Between the
two steps, the transmission $g_2(\Vg)$ has a minimum, corresponding to
the spin gap, and the total conductance $g(\Vg)$ in (d3) likewise
develops a spin gap minimum with increasing $B$. These features can be
understood by looking at the spin composition of the two bulk
dispersion branches, depicted quantitatively through the colors in
Fig.~\ref{fig:disp}(a1-a3). At $k=0$ the SOI field is zero and we have
pure spin-states w.r.t.\ the chosen quantization. At larger $|k|$ the
SOI field increases, leading to spin-mixing. In fact in the limit
$k\rightarrow\infty$ we find a fully mixed state with equal up/down
contributions. Since the upper branch minimum at $k=0$ is in a pure
spin-down state it corresponds to a peak only in
$\mathcal{A}^\downarrow$. But the minima of the lower branch are
shifted away from $k=0$ and have a spin down share besides the
dominant spin up contribution. This causes the emergent peak in
$\mathcal{A}^\downarrow$ at low frequencies, whose height increases
with $R$, due to the stronger spin-mixing.

\textit{Interacting system. }We now include EEI via $H_{\rm
  int}=\sum_j U_j d^\dagger_{j \uparrow} d_{j\uparrow}d^\dagger_{j
  \downarrow} d_{j\downarrow}$. The on-site interaction $U_j = U(ja)$
is switched on smoothly over the QPC according to
$U(x)=U\exp(-(2x/L)^6/(1-(2x/L)^2))$. We set $U_j=0$ for
  $|j|>N$, because outside the QPC region transverse confinement is
  weak or absent, and screening strong
  \cite{bauermethods,BauerNature}. We calculate the conductance at
zero temperature with the functional Renormalization Group technique
in the one-particle irreducible version
\cite{Wetterich1993,birkholzpaper,birkholzthesis,bauerthesis,Metzneretal2012}
using the coupled ladder approximation, which was presented in
\cite{bauermethods} for a model without SOI. Generalizations necessary
in the presence of SOI are described in the Supplement.

\begin{figure}
\includegraphics[width=\columnwidth]{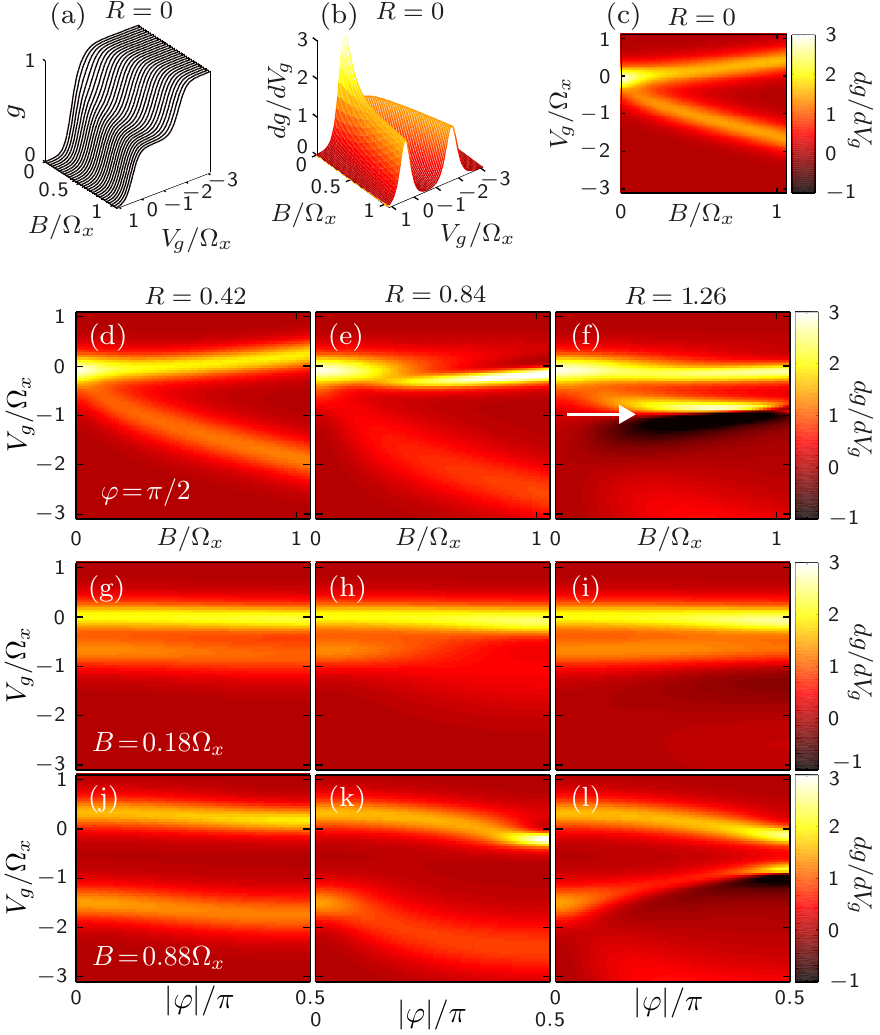}
\caption{\label{fig:colorplots}fRG results for the conductance $g$ and
  transconductance $dg/d\Vg$, for $U=3.36\sqrt{\Omegax\tau}$ at zero
  temperature. Top row: 3d or colorscale plots of the conductance (a)
  and the transconductance (b,c) as functions of $\Vg$ and $B$, for
  $R=0$. Three bottom rows (d-l): Colorscale plots of the
  transconductance for three choices of $R$ (three columns), plotted
  as a function of $\Vg$ and either $B$ for fixed $\varphi = \pi/2$,
  (second row), or of $\varphi$ for fixed $B=0.18\Omegax$ (third row)
  and $B=0.88\Omegax$ (fourth row).  }
\end{figure}
The $B$-dependence of the conductance for $\varphi = \pi/2$ and
different $R$ in the presence of EEI is shown in the right column
(e1-e4) of Fig.~\ref{fig:disp} and the corresponding transconductance
d$g/$d$\Vg$ in Fig.~\ref{fig:colorplots}(b-f). The case $R=0$, see
Figs.~\ref{fig:disp}(e1) and \ref{fig:colorplots}(a-c), which is
equivalent to $\varphi=0$, has been discussed in
\cite{BauerNature,bauermethods}: once a finite magnetic field breaks
the spin degeneracy a surplus of spin-up electrons develops in the
QPC, so that spin-down electrons experience both a Zeeman and a
Coulomb energy cost. This Stoner-type effect depends on the LDOS at
$\mu$ and hence is strongest when the apex of the van Hove ridge
touches the chemical potential, i.e.\ when $\Vg$ is within
$\simeq$$0.5\Omegax$ below 0 \cite{BauerNature}. This causes an
asymmetry w.r.t.\ $\Vg =0$ in the $B$-induced evolution of $g(\Vg)$
from a single to a double step in Fig.~\ref{fig:disp}(e1), in contrast
to the case without EEI in Fig.~\ref{fig:disp}(d1). This asymmetry is
characteristic of the 0.7-anomaly. The corresponding transconductance
in Fig.~\ref{fig:colorplots}(b-c) shows a double peak whose spacing
increases roughly linearly with $B$ (with an EEI-enhanced g-factor),
as seen in numerous experiments
\cite{Thomas1996,Micolich2011,BauerNature}.

The Stoner-type Coulomb enhancement of a field-induced population
imbalance is amplified when $R \neq 0$, as shown in
Figs.~\ref{fig:disp}(e2-e4) and \ref{fig:colorplots}(d-f), because of
the height imbalance for the spin-up and spin-down LDOS peaks caused
by SOI. Correspondingly, with increasing $R$ the double-step structure
in the conductance becomes more pronounced, the second substep
becoming much broader than the first, see Figs.~\ref{fig:disp}(e2-e3),
and the transconductance in Fig.~\ref{fig:colorplots}(d-e) shows a
weakening of the lower-$\Vg$ peak with increasing $R$. This reflects
the increasing curvature $C_{\rm d2}$ of the upper dispersion branch
(and hence larger step width $\Omegaxtwo$). For $R \gtrsim 1 $,
additional features, inherited from the noninteracting case, emerge
for $g(\Vg)$ in Fig.~\ref{fig:disp}(e4): a local maximum (marked by an
arrow), followed by a spin gap minimum at lower $\Vg$. For the
transconductance, Fig.~\ref{fig:colorplots}(f), these features show up
as a strong secondary peak around $\Vg /\Omegax$$\simeq$$-1$ (marked
by an arrow), followed by a region of negative transconductance
(black). EEI also induce a secondary 0.7-type double-step feature in
$g(\Vg)$ for $\Vg/\Omegax$ between $0$ and $-1$,
Fig.~\ref{fig:disp}(e4), which is similar to, but narrower than that
for $R=0$. It originates from the main LDOS peak in
$\mathcal{A}^\uparrow_0$ and the \textit{emergent} peak in
$\mathcal{A}^\downarrow_0$. Unlike the regular
$\mathcal{A}_0^\downarrow$ peak aligned with the upper dispersion
branch, whose $\Vg$-position is governed by the magnetic field, the
emergent $\mathcal{A}_0^\downarrow$ peak occurs, due to strong
spin-mixing, at nearly $B$-independent energy close to the
$\mathcal{A}^\uparrow_0$ peak. As a result, the two transconductance
maxima in Fig.~\ref{fig:colorplots}(f) remain parallel with increasing
$B$, in strong contrast to the situation for $R<1$ in
Fig.~\ref{fig:colorplots}(c-e).
  
Figures~\ref{fig:colorplots}(g-l) show, for two fixed values of $B$,
how the transconductance evolves as $|\varphi|$ is increased from 0 to
$\pi/2$, thus switching on the effects of SOI. The decrease in peak
spacing with increasing $|\varphi|$ in Fig.~\ref{fig:colorplots}(l)
strikingly reflects the increasing importance of spin mixing. The
strong angle-dependence predicted here is a promising candidate for an
experimental test of our theory \cite{prcomm1}.

At small nonzero temperature,  inelastic scattering causes a Fermi-liquid-type reduction of the conductance,
  $g(T,\Vg)/g(0,\Vg) = 1-(T/T_\ast)^2$ for  $T\ll T_\ast$, with a $\Vg$-dependent low-energy scale $T_\ast(V_g)$.  We expect its
  magnitude to be similar to the case without SOI, typically
  $\simeq$1K \cite{BauerNature}.  Thus, for $T \lesssim 0.1$K, the
  $T$-dependence should be very weak and the $T=0$ predictions
  applicable.

In summary, we have shown that in the presence of SOI, the changes in
the dispersion induced by the interplay of $\B$ and $\BSOI$ can
strongly affect the shape of the 0.7-anomaly. In the absence of EEI,
SOI cause an anisotropic response of the spin splitting to the applied
in-plane magnetic field. With EEI, the 0.7-anomaly also develops an
anisotropic response to magnetic field, and if SOI are strong, the
conductance develops additional features due to the interplay of EEI
and SOI: for $\varphi = \pi/2$ these include a field-induced double
step in the conductance that does \textit{not} split linearly with
$B$, followed by a spin gap minimum. The dependence of the conductance
on the angle between $\B$ and $\BSOI$ is already apparent for
$R\simeq0.4$, which is accessible in experiments with electron
QPCs. Hole QPCs with $R \gtrsim 1$ would allow access to regimes with
strong SOI.

An experimental verification of our predictions would highlight
  the influence of LDOS features on the conductance and thus lend
  further support to the van Hove scenario of Ref.~\cite{BauerNature}
  as microscopic explanation for the $0.7$-anomaly. More generally,
  our work lays out a conceptual framework for analyzing the interplay
  of SOI, EEI and barrier shape in quasi-1D geometries: examine how
  SOI and barrier shape modify the (bare) LDOS near $\mu$ -- whenever
  the LDOS is large, EEI effects are strong. We expect
  this to be relevant for the more complicated hybrid
  superconductor-semiconductor junctions currently studied by seekers
  of Majorana fermions \cite{Lutchyn2010,Oreg2010a,Mourik2012}. A proper
  analysis of such systems would require a generalization of our approach to
  include superconducting effects.

We thank S.~Fischer, A.~Hamilton, K.~Hudson, S.~Ludwig,
C.~Marcus, A.~Micolich and A.~Srinivasan for interesting and
useful discussions and acknowledge support from the DFG via SFB-631,
SFB-TR12, De730/4-3, and the Cluster of Excellence \emph{Nanosystems
  Initiative Munich}. O.G. acknowledges support from the NSF under the grant PHY-1314735.

\vspace{-5mm}

\bibliography{soibib}

\clearpage

\appendix

\section*{Supplemental Material to ``Effect of spin-orbit interactions on the 0.7 anomaly in quantum point contacts''}

\subsection{Geometric details of the model}
In our model the 2DES is in the $xy$-plane and the QPC is directed along the $x$-axis (this is the direction of motion of the charge carrier). For the directions of the $\B$ and $\BSOI$ fields we impose the following restrictions. To avoid orbital effects we require the magnetic field $\B$ to be in the $xy$-plane of the 2DES. We also want to be able to rotate $\B$ through any angle $\varphi$ w.r.t.\ $\BSOI$, which implies that $\BSOI$ also must lie in the $xy$-plane. With the latter condition, the $\BSOI$ field can be either parallel to the direction of motion of the electrons (pure Dresselhaus contribution), or orthogonal to it (pure Rashba contribution), or a combination of the two. But for our mathematical model, the end results depend only on the relative angle $\varphi$ between $\B$ and $\BSOI$. This means that we can choose the direction of $\BSOI$ without loss of generality. We choose $\BSOI$ to be parallel to the $y$-axis, c.f.\ Eq.~(2) of the main text.

\subsection{The spatially resolved LDOS}
\begin{figure}
\includegraphics[width = 0.99\columnwidth]{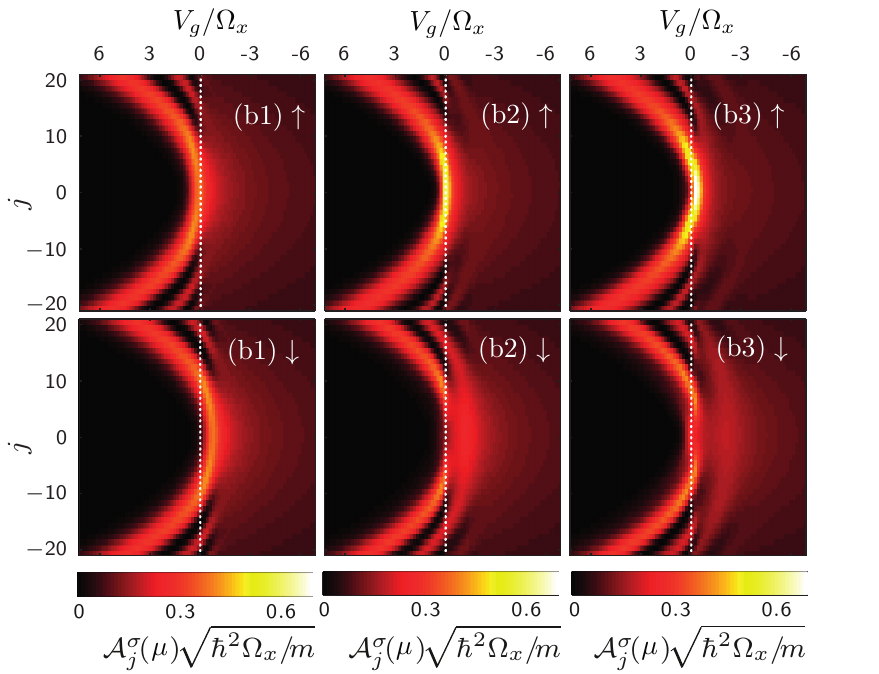}
\caption{\label{fig:supplspatresspec}Spatially resolved plots of the 
noninteracting LDOS $\mathcal{A}_j^\sigma$ at fixed $\omega=\mu$, 
plotted as a function of gate voltage $V_{\rm g}$ and site index $j$, for $B = 
0.88\Omega_x$ and for spin $\sigma = \uparrow$ (top row) and $\sigma = 
\downarrow$ (bottom row). Left column: $R = 0.84$, $\varphi = 0$. Middle column: 
$R = 0.84$, $\varphi = \pi/2$. Right column: $R = 1.26$, $\varphi = \pi/2$. All 
results shown are for $N = 50$.}
\end{figure}

In Fig.~1(b1-b3) of the main text we show the LDOS at fixed $\mu$ as a function 
of $V_{\rm g}$ on the central site of the QPC. The behavior at the center 
captures all relevant features. For completeness we include here in 
Fig.~\ref{fig:supplspatresspec} the spatially resolved plots of the LDOS 
$\mathcal{A}_j^\sigma$ for both spin states and the same parameter values as in 
Fig.~1 of the main text.

\subsection{Second order fRG}
The functional Renormalization Group (fRG) method is an improved perturbation 
technique 
\cite{Wetterich1993,birkholzpaper,birkholzthesis,bauerthesis,Metzneretal2012}. 
Rather than expanding the Green's function in orders of the coupling and 
truncating the expansion, fRG introduces a flow parameter $\Lambda$ into the 
free Green's function $\mathcal{G}_0$. At zero temperature we define
\begin{equation}
\mathcal{G}_0(i\omega)\rightarrow\theta(|\omega|-\Lambda)\mathcal{G}
_0(i\omega)\equiv\mathcal{G}^\Lambda_0(i\omega),
\end{equation}
where $\theta(\omega)$ is the Heaviside step function. At the beginning of the 
flow, setting $\Lambda=\infty$ yields $\mathcal{G}_0^\Lambda=0$, which means 
that the only contribution to the full Green's function comes from the bare 
vertex. At the end of the flow, $\Lambda=0$ recovers the full model. The 
technical details of the one-particle irreducible version of the fRG employed in 
this work are presented in depth in \cite{bauermethods}. We use the static 
approximation at zero temperature, which will be described below in 
Sec.~\ref{sec:approximations}. Since \cite{bauermethods} does not deal with 
spin-orbit interactions, no spin-mixing is possible, which introduces additional 
symmetries into the system. In our case these symmetries are no longer present. 
In this Supplement we focus on the generalizations necessary to account for such 
spin-mixing terms.

The second-order fRG flow equations in the one-particle irreducible version and 
in the static approximation are given by
\begin{widetext}
\begin{eqnarray}
\ddL\gammaone(\color{DGLorange} q_1' \color{black}, \color{RoyalBlue} q^{~}_1
 \color{black}) &=& \frac{1}{2 \pi} \sum_{\omega = \pm \Lambda} 
\sum_{q_2',q_2^{~}}
 \widetilde{\mathcal{G}}_{q^{~}_2,q'_2}^\Lambda (i\omega ) \gammatwo(q_2' 
,\color{DGLorange} q_1' \color{black}; q^{~}_2,\color{RoyalBlue} q^{~}_1  
\color{black}), \label{eq:frg2a} \\
 \ddL\gammatwo &=&\ddL ( \gammap+\gammax+\gammad ),\ \ \textnormal{where}\\
\ddL \gammap(\color{DGLorange} q_1',q_2' \color{black} ;\color{RoyalBlue} 
q^{~}_1,q^{~}_2 \color{black}) 
&=& \frac{1}{2 \pi} \sum_{\omega = \pm \Lambda} \sum_{q_3', q^{~}_3, q_4', 
q^{~}_4}
\frac{1}{2} \gammatwo(\color{DGLorange} q_1',q_2' \color{black} 
;q^{~}_3,q^{~}_4)
\widetilde{\mathcal{G}}^\Lambda_{q^{~}_3, q_3'} (i \omega )
\widetilde{\mathcal{G}}^\Lambda_{q^{~}_4,q_4'} (-i \omega )
\gamma_2^\Lambda (q_3', q_4';\color{RoyalBlue} q^{~}_1,q^{~}_2 \color{black}), 
\\
\ddL\gammax(\color{DGLorange} q_1',q_2' \color{black} ;\color{RoyalBlue} 
q^{~}_1,q^{~}_2 \color{black}) 
&=&\frac{1}{2 \pi} \sum_{\omega = \pm \Lambda} \sum_{q_3', q^{~}_3, q_4', 
q^{~}_4}
\gammatwo(\color{DGLorange} q_1' \color{black} ,q_4';q^{~}_3, \color{RoyalBlue} 
q^{~}_2 \color{black})
\widetilde{\mathcal{G}}^\Lambda_{q^{~}_3,q_3'} (i \omega 
)\widetilde{\mathcal{G}}^\Lambda_{q^{~}_4, q_4'} (i \omega )
\gammatwo(q_3', \color{DGLorange} q_2' \color{black} ;\color{RoyalBlue} q^{~}_1 
\color{black} , q^{~}_4), 
 \\
\ddL  \gammad(\color{DGLorange} q_1',q_2' \color{black} ;\color{RoyalBlue} 
q^{~}_1,q^{~}_2 \color{black}) 
&=&-\frac{1}{2 \pi} \sum_{\omega = \pm \Lambda} \sum_{q_3', q^{~}_3, q_4', 
q^{~}_4}
\gamma_2^\Lambda (\color{DGLorange} q_1' \color{black} ,q_3';\color{RoyalBlue} 
q^{~}_1 \color{black} , q^{~}_4)
\widetilde{\mathcal{G}}^\Lambda_{q^{~}_4 , q_4'} (i \omega 
)\widetilde{\mathcal{G}}^\Lambda_{q^{~}_3,q_3'} (i \omega )
\gammatwo (q_4', \color{DGLorange} q_2' \color{black} ;q^{~}_3, 
\color{RoyalBlue} q^{~}_2 \color{black}), \label{eq:frg2e}
\end{eqnarray}
\end{widetext}
where $-\gammaone$ is the self-energy and $\gammatwo$ is the two-particle 
irreducible vertex. All higher order vertices $\gamma_{n\geq3}$ have been set to 
zero. Here $\widetilde{\mathcal{G}}^\Lambda $ is defined as
\begin{equation}
\widetilde{\mathcal{G}}^\Lambda = \left[\mathcal{G}^{-1}_0 + 
\gamma_1^{\Lambda}\right]^{-1}=\frac{1}{i\omega-H_0+ \gamma_1^{\Lambda}},
\end{equation}
where $H_0$ is the (known) Hamiltonian of the non-interacting system. The 
quantum numbers $q_i$ encode the spin and spatial degrees of freedom 
$q\equiv(\sigma,j)$. The flow of $\gammatwo$ was split into three contributions 
called the particle-particle channel ($P$), and the exchange ($X$) and direct 
($D$) contributions to the particle-hole channel, respectively. This will allow 
us to simplify the flow equations later.

For a derivation of Eqs.~(\ref{eq:frg2a}-\ref{eq:frg2e}) see for instance 
\cite{BauerNature,bauerthesis,bauermethods}. 

\subsubsection{Initial condition}
For the numerical treatment we cannot set the initial value of the flow 
parameter $\Lambda_{\rm init}$ to infinity, but it is sufficient that it is much 
larger than all relevant energy scales. We have the following initial condition 
at $\Lambda_{\rm init}$  \cite{BauerNature,bauerthesis,bauermethods},
\begin{eqnarray}
\gamma_2^{\Lambda_{\rm init}}(q_1,q_2,q_3,q_4)&=&v_{q_1,q_2,q_3,q_4},\\
\gamma_1^{\Lambda_{\rm init}}(q_1,q_2)&=&-\frac{1}{2}\sum_qv_{q_1,q,q_2,q},
\end{eqnarray}
where the vertex $v_{q_1,q_2,q_3,q_4}$ is site diagonal and at site $j\equiv 
j_1=j_2=j_3=j_4$ is given by
\begin{equation}
v_{q_1,q_2,q_3,q_4}=U_j\delta_{\sigma_1\bar{\sigma}_2}\left(\delta_{
\sigma_1\sigma_3}\delta_{\sigma_2\sigma_4}-\delta_{\sigma_1\sigma_4}\delta_{
\sigma_2\sigma_3}\right).
\label{eq:initgammatwo}
\end{equation}
This means that the spins $q_1$ and $q_2$, as well as the spins $q_3$ and $q_4$ 
must be opposite. This leaves two possibilities: 
$\sigma_1=\sigma_3=\bar{\sigma}_2=\bar{\sigma}_4$ that has positive sign, and 
$\sigma_1=\sigma_4=\bar{\sigma}_2=\bar{\sigma}_3$ that has negative sign. 
Inserting this into the initial condition for $\gammaone$ yields
\begin{equation}
\gamma_1^{\Lambda_{\rm init}}(q',q)=-(U_j/2)\delta_{\sigma'\sigma}.
\end{equation}

\subsubsection{Approximations}
\label{sec:approximations}
We use the following approximations, see \cite{BauerNature} and references 
thereof. Firstly, we neglect the frequency dependence of $\gammatwo$. This is 
called the static approximation and is known to give good results at $T=0$ 
\cite{bauermethods}. Given the structure of the flow equation for $\gammatwo$ 
above, it is natural to divide the flowing vertex into four parts as follows:
\begin{equation}
\label{eq:gammatwo}\gammatwo  = v + \gammap + \gammax + \gammad.
\end{equation}
Here $v$ is shorthand for the bare vertex, and the flows of $\gammap$, $\gammax$ 
and $\gammad$ were given above. The initial value for $\gammatwo$ is simply the 
bare vertex. If we insert the bare vertex into the flow equations for the 
components of $\gammatwo$ we observe that some of its symmetries remain 
preserved in the derivative on the left hand side. For instance, for $\gammap$ 
we see that the first two and the second two site indices must be identical and 
also that the first and the second pair of spin indices must be opposite, 
respectively. Similarly, for $\gammax$ we see that the first and fourth site 
index, as well as the the second and third site index must be equal. For 
$\gammad$ the first and the third, as well as the second and the fourth site 
indices are equal. There is no restriction on the any of the spin indices for 
either $\gammax$ or $\gammad$.

The next iteration would break the remaining symmetries, since all three 
channels contribute to the derivative of $\gammatwo$ and then back-feed into the 
differential equations for each channel. If instead we choose to only back-feed 
each channel into its own differential equation, we can preserve the symmetries 
described above. This immensely simplifies the treatment of the equations. By 
doing so, we only neglect contributions of order $v^3$ and higher, which 
justifies their neglect as long as $U_j$ is not too large. Altogether we obtain 
the following contributions:
\begin{eqnarray}
P^{\sigma\bar{\sigma}}_{ji} &:=&
\gammap\left(\color{DGLorange} j\sigma,j \bar{\sigma}\color{black} ; 
\color{RoyalBlue} i\sigma , i \bar{\sigma}\color{black} \right) \\
\bar{P}^{\sigma\bar{\sigma}}_{ji} &:=&
\gammap\left(\color{DGLorange} j\sigma,j \bar{\sigma}\color{black} ; 
\color{RoyalBlue} i\bar{\sigma} , i \sigma\color{black} \right) \\
X^{\sigma_1\sigma_2\sigma_3\sigma_4}_{ji}  &:=&
\gammax \left( \color{DGLorange} j\sigma_1 , i \sigma_2\color{black} ; 
\color{RoyalBlue} i\sigma_3 , j \sigma_4\color{black} \right)\\
D^{\sigma_1\sigma_2\sigma_3\sigma_4}_{ji} &:=&
\gammad \left(\color{DGLorange} j\sigma_1 , i \sigma_2\color{black} ; 
\color{RoyalBlue} j\sigma_3 , i \sigma_4 \color{black} \right)
\end{eqnarray}
Note that some elements of a channel can also satisfy the symmetries of another 
channel. So for instance the diagonal element $P^{\sigma\bar{\sigma}}_{jj}$ has 
the same symmetries as the diagonal elements 
$X^{\sigma\bsigma\sigma\bsigma}_{jj}$ and $D^{\sigma\bsigma\sigma\bsigma}_{jj}$. 
If we back-feed such elements too we preserve the symmetries in each channel, 
but obtain a more accurate approximation. Therefore, in each of the three flow 
equations for the channels of $\gammatwo$ we replace $\gammatwo$ on the right 
hand side by the appropriate channels plus the site diagonal contributions of 
the other channels that obey the same symmetries. The initial conditions for the 
three channels follow immediately from (\ref{eq:gammatwo}): 
$\gamma_p^{\Lambda_{\rm init}}=\gamma_x^{\Lambda_{\rm 
init}}=\gamma_d^{\Lambda_{\rm init}}=0$. Of course, for the differential 
equation for $\gammaone$ we need the full $\gammatwo$ which is the sum of all 
three channels and the bare vertex.

\subsubsection{Symmetries}
Due to the hermiticity of the Hamiltonian the following relation holds for the 
Green's function
\begin{equation}
\mathcal{G}(i\omega)=\mathcal{G}^\dagger(-i\omega)\Leftrightarrow 
\mathcal{G}_{ij}(i\omega)=\mathcal{G}^*_{ji}(-i\omega).
\label{eq:gfunsym}
\end{equation}
We assume that this relation also holds for $\tilde{\mathcal{G}}^\Lambda$. If 
$\gammaone$ is hermitian then the assumption is obviously justified. Numerical 
results indeed confirm that $\gammaone$ is hermitian. We also have the following 
symmetries of $\gamma_2$:
\begin{eqnarray}
\gamma_2(\color{DGLorange}q_1,q_2,\color{RoyalBlue}q_3,q_4\color{black}
)&=&-\gamma_2(\color{DGLorange}q_2,q_1\color{black},\color{RoyalBlue}q_3,
q_4\color{black})\label{eq:symgamma1}\\
&=&-\gamma_2(\color{DGLorange}q_1,q_2\color{black},\color{RoyalBlue}q_4,
q_3\color{black})\label{eq:symgamma2}\\
&=&\gamma_2(\color{DGLorange}q_2,q_1\color{black},\color{RoyalBlue}q_4,q_3\color
{black})\label{eq:symgamma3}
\end{eqnarray}
This follows directly from the equation defining the two-particle vertex, see 
e.g.\ \cite{bauerthesis}. Again we assume that these relations hold also for 
$\gammatwo$ and moreover for each of the separate channels. Their consistency 
with the numerical results will be demonstrated below. Altogether this yields 
the following symmetry relations for the different channels:
\begin{eqnarray}
& 
&P^{\sigma\bar{\sigma}}_{ji}=P^{\bar{\sigma}\sigma}_{ji}=-\bar{P}^{\sigma\bar{
\sigma}}_{ji}\\
& 
&D_{ji}^{\sigma_1\sigma_2\sigma_3\sigma_4}=D_{ij}^{
\sigma_2\sigma_1\sigma_4\sigma_3}=-X^{\sigma_1\sigma_2\sigma_4\sigma_3}_{ji}\\
& 
&X^{\sigma_1\sigma_2\sigma_3\sigma_4}_{ji}=X^{\sigma_2\sigma_1\sigma_4\sigma_3}_
{ij}=-D^{\sigma_1\sigma_2\sigma_4\sigma_3}_{ji}
\end{eqnarray}
We observe that $P^{\uparrow\downarrow}=P^{\downarrow\uparrow}$ and hence the 
spin indices for $P$ will be dropped from now on, leaving only the site index. 
The alternative configuration $\bar{P}$ follows completely from $P$ and does not 
need to be kept track of separately. Same applies to $X$ and $D$ which 
completely define each other. We choose to work with $D$. There are various 
symmetries of $D$ but there is no restriction on the spin index. This means that 
there are $2^4=16$ different submatrices corresponding to $16$ different spin 
configurations of $D$. We choose to arrange them as follows
\begin{equation}
D^{\color{RoyalBlue}\sigma_1\color{black}\sigma_2\color{RoyalBlue}\sigma_3\color
{black}\sigma_4}=\left(\begin{array}{c|c|c|c}
\color{RoyalBlue}\uparrow\color{black}\uparrow\color{RoyalBlue}\uparrow\color{black}\uparrow & 
\color{RoyalBlue}\uparrow\color{black}\uparrow\color{RoyalBlue}\uparrow\color{black}\downarrow & 
\color{RoyalBlue}\uparrow\color{black}\downarrow\color{RoyalBlue}\uparrow\color{black}\uparrow & 
\color{RoyalBlue}\uparrow\color{black}\downarrow\color{RoyalBlue}\uparrow\color{black}\downarrow \\ \hline
\color{RoyalBlue}\uparrow\color{black}\uparrow\color{RoyalBlue}\downarrow\color{black}\uparrow & 
\color{RoyalBlue}\uparrow\color{black}\uparrow\color{RoyalBlue}\downarrow\color{black}\downarrow & 
\color{RoyalBlue}\uparrow\color{black}\downarrow\color{RoyalBlue}
\downarrow\color{black}\uparrow & 
\color{RoyalBlue}\uparrow\color{black}\downarrow\color{RoyalBlue}
\downarrow\color{black}\downarrow \\ \hline
\color{RoyalBlue}\downarrow\color{black}\uparrow\color{RoyalBlue}\uparrow\color{black}\uparrow & 
\color{RoyalBlue}\downarrow\color{black}\uparrow\color{RoyalBlue}\uparrow\color{black}\downarrow & 
\color{RoyalBlue}\downarrow\color{black}\downarrow\color{RoyalBlue}
\uparrow\color{black}\uparrow & 
\color{RoyalBlue}\downarrow\color{black}\downarrow\color{RoyalBlue}
\uparrow\color{black}\downarrow \\ \hline
\color{RoyalBlue}\downarrow\color{black}\uparrow\color{RoyalBlue}
\downarrow\color{black}\uparrow & 
\color{RoyalBlue}\downarrow\color{black}\uparrow\color{RoyalBlue}
\downarrow\color{black}\downarrow & 
\color{RoyalBlue}\downarrow\color{black}\downarrow\color{RoyalBlue}
\downarrow\color{black}\uparrow & 
\color{RoyalBlue}\downarrow\color{black}\downarrow\color{RoyalBlue}
\downarrow\color{black}\downarrow
\end{array}\right)\label{eq:spinmatrixscheme}
\vspace{0.1mm}
\end{equation}
Note that the first and third spin index are fixed along a row and 
correspondingly the second and fourth index are fixed along one column. This 
form of the matrix will prove convenient later. From the symmetries of $D$ it 
follows that this matrix is symmetric. Numerically we also confirm the following 
relations between the different blocks, schematically
\[\begin{array}{|c|c|c|c|} \hline A_s & B & B^* & C \\ \hline
B^T & D_s & E_h & F \\ \hline
(B^*)^T & E_h^T & D_s^* & F^* \\ \hline
C^T & F^T & (F^*)^T & G_s \\ \hline
\end{array}\]
where identical symbols denote equal blocks and symmetric (hermitian) 
submatrices are labeled by the subscript $s$ ($h$). There are only seven 
different blocks in total. Numerically we also show that the corner submatrices 
$A_s$, $G_s$, $C$ and $C^T$ are real. The other submatrices are complex in 
general. For a hermitian $\gammaone$, the first flow equation implies that 
$\gammatwo(q_2' ,q_1' ; q_2, q_1)= \gamma_2^{\Lambda *}(q_2 ,q_1 ; q'_2, q'_1)$. 
Translated to the separate channels this confirms that $P$ must indeed be 
hermitian, since $P_{ij}=P_{ji}^*$, as well as all the remaining relations 
between the different submatrices of $D$.

\begin{widetext}
\subsubsection{Flow equation for the P-channel}
Restricting $\gammap$ according to the symmetries of the $P$-channel we obtain 
the following simplified equation for the derivative of $P$:
\begin{eqnarray}
\ddL P_{ji}&=&\ddL\gammap\left(\color{DGLorange} j\sigma,j 
\bar{\sigma}\color{black} ; \color{RoyalBlue} i\sigma , i 
\bar{\sigma}\color{black} \right) \\
&=& \frac{1}{2 \pi} \sum_{\omega = \pm \Lambda} \sum_{k,l}\frac{1}{2}\cdot\left[
\gammatwo(\color{DGLorange} j\sigma,j \bar{\sigma} \color{black} 
;k\sigma,k\bar{\sigma})
\widetilde{\mathcal{G}}^{\Lambda\sigma\sigma}_{kl} (i \omega )
\widetilde{\mathcal{G}}^{\Lambda\bsigma\bsigma}_{kl} (-i \omega )
\gammatwo (l\sigma,l\bar{\sigma};\color{RoyalBlue}  i\sigma , i \bar{\sigma} 
\color{black})\right. \nonumber\\
&~&\hspace{3cm}+\gammatwo(\color{DGLorange} j\sigma,j \bar{\sigma} \color{black} 
;k\bsigma,k\sigma)
\widetilde{\mathcal{G}}^{\Lambda\bsigma\bsigma}_{kl} (i \omega )
\widetilde{\mathcal{G}}^{\Lambda\sigma\sigma}_{kl} (-i \omega )
\gammatwo (l\bsigma,l\sigma;\color{RoyalBlue}  i\sigma , i \bar{\sigma} 
\color{black}) \nonumber\\
&~&\hspace{3cm}+\gammatwo(\color{DGLorange} j\sigma,j \bar{\sigma} \color{black} 
;k\sigma,k\bar{\sigma})
\widetilde{\mathcal{G}}^{\Lambda\sigma\bsigma}_{kl} (i \omega )
\widetilde{\mathcal{G}}^{\Lambda\bsigma\sigma}_{kl} (-i \omega )
\gammatwo (l\bsigma,l\sigma;\color{RoyalBlue}  i\sigma , i \bar{\sigma} 
\color{black}) \nonumber\\
&~&\hspace{3cm}\left.+\gammatwo(\color{DGLorange} j\sigma,j \bar{\sigma} 
\color{black} ;k\bsigma,k\sigma)
\widetilde{\mathcal{G}}^{\Lambda\bsigma\sigma}_{kl} (i \omega )
\widetilde{\mathcal{G}}^{\Lambda\sigma\bsigma}_{kl} (-i \omega )
\gammatwo (l\sigma,l\bar{\sigma};\color{RoyalBlue}  i\sigma , i \bar{\sigma} 
\color{black})\right]
\end{eqnarray}
Note that the first two terms and the last two terms in the sum are equivalent 
after summation over $\omega$, due to the symmetry relations 
(\ref{eq:symgamma2}) and (\ref{eq:symgamma3}). We can thus keep one of the terms 
respectively and cancel the factor of $1/2$. With the definitions
\begin{eqnarray}
\Pi_{kl}^{p\Lambda(1)}&=&\frac{1}{2 \pi} \sum_{\omega = \pm 
\Lambda}\widetilde{\mathcal{G}}^{\Lambda\bsigma\bsigma}_{kl} (i \omega )
\widetilde{\mathcal{G}}^{\Lambda\sigma\sigma}_{kl} (-i \omega )\\
\Pi_{kl}^{p\Lambda(2)}&=&\frac{1}{2 \pi} \sum_{\omega = \pm 
\Lambda}\widetilde{\mathcal{G}}^{\Lambda\bsigma\sigma}_{kl} (i \omega )
\widetilde{\mathcal{G}}^{\Lambda\sigma\bsigma}_{kl} (-i \omega )
\end{eqnarray}
the flow equation can be written more succinctly as
\begin{eqnarray}
\ddL P_{ji}&=&\sum_{kl}\left[\gammatwo(\color{DGLorange} j\sigma,j \bar{\sigma} 
\color{black} ;k\sigma,k\bar{\sigma})\Pi_{kl}^{p\Lambda(1)}\gammatwo 
(l\sigma,l\bar{\sigma};\color{RoyalBlue}  i\sigma , i \bar{\sigma} 
\color{black})+\gammatwo(\color{DGLorange} j\sigma,j \bar{\sigma} \color{black} 
;k\sigma,k\bar{\sigma})\Pi_{kl}^{p\Lambda(2)}\gammatwo 
(l\bsigma,l\sigma;\color{RoyalBlue}  i\sigma , i \bar{\sigma} 
\color{black})\right]\\
&=&\sum_{kl}\gammatwo(\color{DGLorange} j\sigma,j \bar{\sigma} \color{black} 
;k\sigma,k\bar{\sigma})\left[\Pi_{kl}^{p\Lambda(1)}-\Pi_{kl}^{p\Lambda(2)}\right
]\gammatwo (l\sigma,l\bar{\sigma};\color{RoyalBlue}  i\sigma , i \bar{\sigma} 
\color{black}),
\end{eqnarray}
where in the last step we used symmetry relation (\ref{eq:symgamma2}). If we now 
define
\begin{equation}
\Pi_{kl}^{p\Lambda}\equiv\Pi_{kl}^{p\Lambda(1)}-\Pi_{kl}^{p\Lambda(2)}=\frac{1}{
2 \pi} \sum_{\omega = \pm 
\Lambda}\left[\widetilde{\mathcal{G}}^{\Lambda\bsigma\bsigma}_{kl} (i \omega )
\widetilde{\mathcal{G}}^{\Lambda\sigma\sigma}_{kl} (-i \omega 
)-\widetilde{\mathcal{G}}^{\Lambda\bsigma\sigma}_{kl} (i \omega )
\widetilde{\mathcal{G}}^{\Lambda\sigma\bsigma}_{kl} (-i \omega )\right]
\end{equation}
we arrive at
\begin{equation}
\ddL P_{ji}=\tilde{P}_{jk}\Pi_{kl}^{p\Lambda}\tilde{P}_{li},
\end{equation}
where $\tilde{P}$ equals $P$ plus the diagonal contributions from the other 
channels which have the same symmetries as $P$. Explicitly we get
\begin{equation}
\tilde{P}_{jk}={P}_{jk}+\delta_{jk}\left(X^{\sigma\bsigma\sigma\bsigma}_{jj}+D^{\sigma\bsigma\sigma\bsigma}_{jj}+U_j\right)
={P}_{jk}+\delta_{jk}\left(-D^{\sigma\bsigma\bsigma\sigma}_{jj}+D^{\sigma\bsigma\sigma\bsigma}_{jj}+U_j\right)
\end{equation}
Note also that the matrix $\Pi_{kl}^{p\Lambda}$ is hermitian, due to the 
symmetry (\ref{eq:gfunsym}) of the Green's function.

\subsubsection{Flow equation for the D-channel}
Restricting $\gammad$ according to the symmetries of the $D$-channel we obtain 
the following simplified equation for the derivative of $D$:
\begin{eqnarray}
\ddL D^{\sigma_1\sigma_2\sigma_3\sigma_4}_{ji}&=&\ddL  \gammad 
\left(\color{DGLorange} j\sigma_1 , i \sigma_2\color{black} ; \color{RoyalBlue} 
j\sigma_3 , i \sigma_4 \color{black} \right) \\
&=&-\frac{1}{2 \pi} \sum_{\omega = \pm \Lambda} 
\sum_{kl}\sum_{\sigma,\sigma',\sigma'',\sigma'''}
\gammatwo(\color{DGLorange} j\sigma_1 \color{black} ,k\sigma;\color{RoyalBlue} 
j\sigma_3 \color{black} , k\sigma')
\widetilde{\mathcal{G}}^{\Lambda\sigma'\sigma''}_{kl} (i \omega 
)\widetilde{\mathcal{G}}^{\Lambda\sigma'''\sigma}_{lk} (i \omega )
\gammatwo (l\sigma'', \color{DGLorange} i\sigma_2 \color{black} ;l\sigma''', 
\color{RoyalBlue} i\sigma_4 \color{black})
\end{eqnarray}
Observe that the summation goes over the second and fourth index of the first 
$\gammatwo$ matrix and over the first and third index of the second $\gammatwo$ 
matrix, while the other indices remain fixed. If we want to recast this 
expression as a matrix multiplication this indeed implies that the first and 
third spin index should be fixed along a row and the second and fourth index 
along one column. This justifies the matrix scheme (\ref{eq:spinmatrixscheme}). 
If we arrange the spin configurations according to this scheme we obtain the 
matrix equation
\begin{equation}
\ddL 
D^{\sigma_1\sigma_2\sigma_3\sigma_4}_{ji}=\sum_{kl}\sum_{\sigma,\sigma',\sigma''
,\sigma'''}
\gammatwo(\color{DGLorange} j\sigma_1 \color{black} ,k\sigma;\color{RoyalBlue} 
j\sigma_3 \color{black} , k\sigma')
\Pi^{d\Lambda\sigma\sigma''\sigma'\sigma'''}_{kl}
\gammatwo (l\sigma'', \color{DGLorange} i\sigma_2 \color{black} ;l\sigma''', 
\color{RoyalBlue} i\sigma_4 \color{black})
\end{equation}
where
\begin{equation}
\Pi^{d\Lambda\sigma\sigma''\sigma'\sigma'''}_{kl}\equiv-\frac{1}{2 \pi} 
\sum_{\omega = \pm \Lambda}\widetilde{\mathcal{G}}^{\Lambda\sigma'\sigma''}_{kl} 
(i \omega )\widetilde{\mathcal{G}}^{\Lambda\sigma'''\sigma}_{lk} (i \omega 
).\label{eq:Pid}
\end{equation}
Note that the order of the spin indices on $\Pi$ is not the same as on the 
Green's functions. The symmetries from (\ref{eq:spinmatrixscheme}) remain valid. 
With our approximation we get
\begin{equation}\ddL 
D^{\sigma_1\sigma_2\sigma_3\sigma_4}_{ji}=\tilde{D}^{\sigma_1\sigma\sigma_3\sigma'}_{jk}\Pi^{d\sigma\sigma''\sigma'\sigma'''}_{kl}
\tilde{D}^{\sigma''\sigma_2\sigma'''\sigma_4}_{li}\end{equation}
where $\tilde{D}$ equals $D$ plus the diagonal contributions from the other 
channels which have the same symmetries as $D$. Explicitly we get
\begin{eqnarray}
\tilde{D}^{\sigma_1\sigma_2\sigma_3\sigma_4}_{jk}&=&{D}^{\sigma_1\sigma_2\sigma_3\sigma_4}_{jk}+\delta_{jk}\left(X^{\sigma_1\sigma_2\sigma_3\sigma_4}_{jj}+(P_{jj}+U_j)\delta_{\sigma_1\bsigma_2}
(\delta_{\sigma_1\sigma_3}\delta_{\sigma_2\sigma_4}-\delta_{\sigma_1\sigma_4}
\delta_{\sigma_2\sigma_3})\right)\\
&=&{D}^{\sigma_1\sigma_2\sigma_3\sigma_4}_{jk}+\delta_{jk}\left(-D^{\sigma_1\sigma_2\sigma_4\sigma_3}_{jj}+(P_{jj}+U_j)\delta_{\sigma_1\bsigma_2}
(\delta_{\sigma_1\sigma_3}\delta_{\sigma_2\sigma_4}-\delta_{\sigma_1\sigma_4}
\delta_{\sigma_2\sigma_3})\right)
\end{eqnarray}
Just like $D$ itself, the matrix 
$\Pi^{d\Lambda\sigma\sigma''\sigma'\sigma'''}_{kl}$ is symmetric, however in 
general not real. The structure of 
$\Pi^{d\Lambda\sigma\sigma''\sigma'\sigma'''}_{kl}$ in terms of its submatrices 
is the same as for $D$.

\subsubsection{Flow equation for $\gamma_1$}
For the self-energy equation
\begin{equation}\ddL\gammaone(\color{DGLorange} k'\sigma' \color{black}, 
\color{RoyalBlue} k\sigma \color{black})
= \frac{1}{2 \pi} \sum_{\omega = \pm \Lambda} \sum_{k_1,k_2,\sigma_1,\sigma_2}
 \widetilde{\mathcal{G}}_{k_2k_1}^{\Lambda\sigma_2\sigma_1} (i\omega ) 
\gammatwo(k_1\sigma_1 ,\color{DGLorange} k'\sigma' \color{black}; 
k_2\sigma_2,\color{RoyalBlue} k\sigma\color{black})\end{equation}
we need the full $\gammatwo=v+\gammap+\gammax+\gammad$. We abbreviate
\begin{equation}\mathcal{S}_{k_2k_1}^{\sigma_2\sigma_1}=\frac{1}{2 \pi} 
\sum_{\omega = \pm 
\Lambda}\widetilde{\mathcal{G}}_{k_2k_1}^{\Lambda\sigma_2\sigma_1} (i\omega 
).\end{equation}
Taking into account the symmetry of each channel we obtain
\begin{eqnarray}
\ddL\gammaone(k'\sigma' ,  k\sigma) 
&=&\delta_{\sigma\sigma'}\mathcal{S}^{\bsigma\bsigma}_{kk'}(P_{k'k}+\delta_{kk'}
U_k)-\delta_{\sigma\bsigma'}\mathcal{S}^{\bsigma\sigma}_{kk'}(P_{k'k}+\delta_{kk'}U_k)\nonumber\\
& 
&-\sum_{\sigma_1\sigma_2}\mathcal{S}^{\sigma_2\sigma_1}_{k'k}D^{\sigma_1\sigma'\sigma\sigma_2}_{kk'}+\delta_{kk'}\sum_{l,\sigma_1,\sigma_2}
\mathcal{S}^{\sigma_2\sigma_1}_{ll}D^{\sigma_1\sigma'\sigma_2\sigma}_{lk}.
\end{eqnarray}
The first line accounts for the bare vertex and the $P/\bar{P}$-channel, while 
the second line contains the contribution from the $X$-channel and then the 
$D$-channel. Note that the $D$-channel only influences the diagonal elements of 
$\gammaone$, due to its symmetry.
\end{widetext}

\end{document}